\documentclass[final]{aipproc}
\usepackage{epsfig}
\usepackage{amssymb}
\layoutstyle{6x9}

\begin{document}

\title[]{Some thoughts on di-jet correlation in Au + Au collisions from PHENIX}

 \keywords{}

\author{Jiangyong Jia for the PHENIX Collaboration}{
  address={Columbia University, New York, NY 10027 and Nevis Laboratories, Irvington, NY 10533, USA}
}
\begin{abstract}
PHENIX has measured the two particle azimuth correlation in Au +
Au at $\sqrt{s}$ = 200 GeV. Jet shape and yield at the away side
are found to be strongly modified at intermediate and low $p_T$,
and the modifications vary dramatically with $p_T$ and centrality.
At high $p_T$, away side jet peak reappears but the yield is
suppressed. We discuss the possible physics pictures leading to
these complicated modifications.
\end{abstract}

\maketitle


High $p_T$ back-to-back jets are valuable probes for the
sQGP~\cite{RHIC} created in heavy-ion collisions at RHIC. To date,
the interactions of jets with the medium are studied via two
particle azimuth correlation method, where particles from one
momentum range (triggering particles) are correlated with those
from another momentum range (associated particles).

In two particle correlation method, the jet multiplicity is
directly related to the number of associated pairs per trigger or
per-trigger yield, $Y =
\frac{1}{{N_{\rm{trigs}}}}\frac{{dN}}{{d\Delta \phi }}$. And the
medium modifications of jet/dijet multiplicity are typically
quantified by $I_{\rm{AA}}$, the ratio of the per-trigger yield in
A + A collisions to that in p + p collisions, $I_{\rm{AA}}  =
\frac{{Y_{\rm{AA}} }}{{Y_{\rm{pp}} }}$. Since the single particle
yield, thus the number of triggering particles, is suppressed in A
+ A collisions, the per-trigger yield defined this way contains a
trivial $R_{\rm{AA}}$ factor. We can reach the following simple
relation between the per-trigger yield using the high $p_T$
particles (denoted as `a') as triggers and that using low $p_T$
particles (denoted as `b') as triggers:
\begin{eqnarray}
\label{eq:eq1} I_{\rm{AA}}^a R_{\rm{AA}}^a  = I_{\rm{AA}}^b
R_{\rm{AA}}^b = \frac{{\rm{JetPairs}_{\rm{AA}} }}{{N_{\rm{coll}}
\times \rm{JetPairs}_{\rm{pp}} }} \equiv J_{\rm{AA}}(p_T^a,p_T^b)
\end{eqnarray}
Where JetPairs$_{\rm{AA}}$ and JetPairs$_{\rm{pp}}$ represent the
average number of jet pairs in one A + A collision and one p + p
collision, respectively. In addition, we define $J_{\rm{AA}} =
\frac{{\rm{JetPairs}_{\rm{AA}} }}{{N_{\rm{coll}}  \times
\rm{JetPairs}_{\rm{pp}} }}$ as the nuclear modification factor of
jet pairs. In energy loss picture, surface bias leads to a
stronger suppression of away side jet yield than that for single
particle yield: $I_{\rm{AA}}<R_{\rm{AA}}$~\cite{Drees:2003zh},
thus $J_{\rm{AA}}< R^a_{\rm{AA}}R^b_{\rm{AA}}$. However, the
preliminary results of the di-jet correlation at high $p_T$ from
STAR shows that $J_{\rm{AA}}\approx
R^a_{\rm{AA}}R^b_{\rm{AA}}$~\cite{Magestro:2005vm}.

The distinctions between the triggering particles and associated
particles are arbitrary. $I_{\rm{AA}}^a$ and $I_{\rm{AA}}^b$ are
not independent of each other as shown in Eq.\ref{eq:eq1}. For
simplicity, we shall define triggering particles as those from the
higher $p_T$ range in following discussions.

Jet correlations at different $p_T$ reflect different aspect of
the interaction between jet and the medium. Previous results
indicate a seemly complete disappearance of the away side jet
signal at moderately high associated $p_T$ (2-4 GeV/c)
~\cite{Adler:2002tq}.
\begin{figure}[t]
\epsfig{file=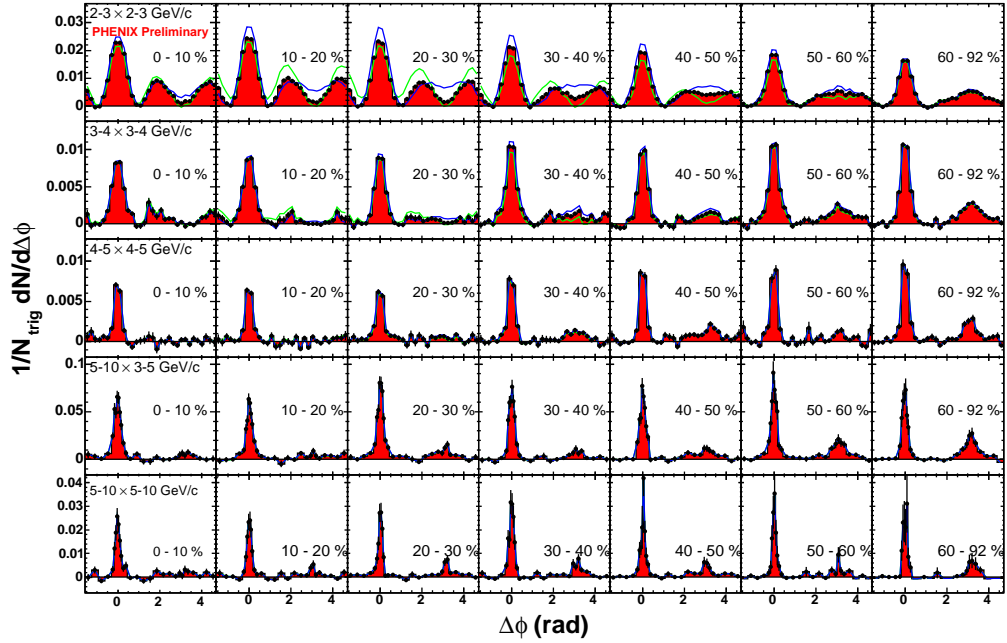,width=0.9\linewidth}
\caption{\label{fig:scan} Background subtracted per-trigger jet
yield in $\Delta\phi$ as function of $p_T$ (vertical) and
centrality (horizontal).}
\end{figure}
At low associated hadron $p_T$, an enhancement of the away side
jet yield~\cite{Adams:2005ph} and a broadened jet
shape~\cite{Adler:2005ee} were observed. These observations are
qualitatively consistent with the energy loss picture, where the
high $p_T$ jets are quenched by the medium and their lost energy
enhanced the jet multiplicity at low $p_T$. Recent results from
PHENIX~\cite{Jia:2005ab} and STAR~\cite{Magestro:2005vm} covering
broader $p_T$ ranges of the triggering and associated particles
provide additional important handles for understanding the
detailed interplay between various competing mechanisms.
Fig.\ref{fig:scan} shows a summary plot of the per-trigger yield
as function of both $p_T$ (vertical) and centrality (horizontal).
Along the vertical direction, we can see how the away side jet
evolves from a cone type of structure at intermediate $p_T$ (2-4
GeV/c) to a relatively flat distribution at moderately-high $p_T$
(3-5 GeV/c), to a reappeared jet structure at high $p_T$ (5-10
GeV/c)~\footnote{The di-jet modification pattern depends on the
$p_T$ of both particles, thus in principle, we should have fixed
the triggering particle $p_T$ and vary the associated particle
$p_T$. However, the $p_T$ selections in Fig.\ref{fig:scan} are
somewhat mixed up in favor of better statistical accuracy.}. The
cone structure qualitatively agree with the `mach cone'/`shock
wave' mechanism \cite{Casalderrey-Solana:2004qm}, which represents
the collective excitation of the medium by the supersonic partons
traversing the medium. The peak structure at the highest $p_T$ bin
could be explained by the tangential contribution of the away side
jet, when both jets are emitted tangential to the surface. The
magnitude of this the peak is reduced compare to the peripheral
collisions, reflecting the reduction in the available phase space
due to the tangential emission requirement. The tangential
contribution must be suppressed even more at lower $p_T$, in order
to describe the relatively flat distribution at the away side.
Along the horizontal direction, the away side jet modifications in
various $p_T$ ranges depend strongly on centrality, consistent
with a smooth turn on of various medium effects from peripheral to
central collisions. Further detailed discussions can be found
in~\cite{Jia:2005ab}.

\begin{figure}[t]
\epsfig{file=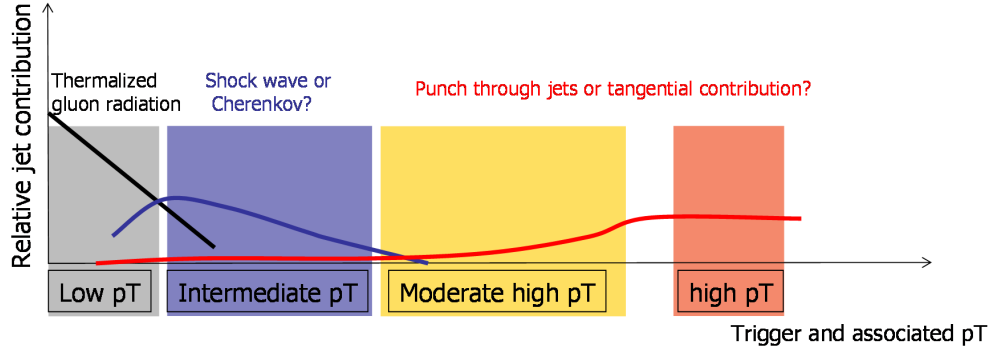,width=0.9\linewidth}
\caption{\label{fig:pic} A sketch of various contributions to the
away side jet yield as function of associated particle and
triggering particle $p_T$. The $p_T$ axis is schematically divided
into four different regions. It is chosen such that the
$p_{T,\rm{trig}}\geq p_{T,\rm{assoc}}$ (see Fig.\ref{fig:scan})).}
\end{figure}

In Fig.\ref{fig:pic}, we draw a crude picture based on the current
results. Different mechanisms play different roles at different
$p_T$. Generally speaking, at low associated particle $p_{T}$, the
jet yield is dominated by thermalized gluon radiation and this
contribution should have a typical exponential thermal spectra and
dies out quickly at $p_T>1-2$ GeV/c; at intermediate $p_T$ (1-4
GeV/c), various collective excitation modes of the medium could
become important. In the case of Mach cone, it contribution to the
yield has an approximate exponential shape; there is a
punching-through component, representing the fragmentation from
the surviving primary jets. Since the away side jet distribution
is very flat at moderately-high $p_T$ (2-5 GeV/c), the
punching-through contribution in this and lower $p_T$ bin has to
be very small; at high $p_T$ ($>3-5$ GeV/c), tangential
contributions could be important and responsible for the clear
away side peak structure in central Au + Au collisions.

\end{document}